\documentstyle[nato,numreferences,epsf,graphicx]{crckapb}
\newcommand{\ampm}{A_{\mu}^{\pm}}
\newcommand{\ampmp}{A_{\mu}^{\pm\,\prime}}
\newcommand{\ampmpp}{A_{\mu}^{\pm\,\prime\prime}}
\newcommand{\amth}{A_{\mu}^{3}}

\newcommand{\be}{\begin{equation}}
\newcommand{\ee}{\end{equation}}

\newcommand{\gmag}{$G_{mag}\,$}

\begin{opening}
\title{CONFINEMENT IN SU(3):\protect\\
       SIMPLE AND GENERALIZED MAXIMAL ABELIAN GAUGE}

\author{John D. Stack and William W. Tucker}
\institute{Department of Physics\\
University of Illinois at Urbana-Champaign\\
1110 W. Green St.\\
Urbana, IL 61801
}

\author{Roy J. Wensley}
\institute{Department of Physics and Astronomy\\
           Saint Mary's College\\
Moraga,CA 94575
}

\end{opening}

\begin{document}


\section{Introduction}
Almost all lattice work on confinement has been carried out for an
$SU(2)$ gauge group.  This is a good starting point.
Confinement is caused by glue and
every physicist believes that there is no essential physical difference
in the confining properties of glue for $SU(N)$, for arbitrary $N$.
Despite this, in confinement  by topological objects, questions 
arise for $SU(3)$ which have no $SU(2)$ analog.
We discuss two of these.  The first involves the fact that the
formulation of the maximal abelian gauge (MAG) is more subtle for
$SU(3)$ than it is for $SU(2)$.  Calculations with the simplest
form give poor results for the string tension (Sec.(\ref{sec_mag})).  
A generalized form which
appears more natural is suggested (Sec.(\ref{sec_mag_gen})).
The second question has to do with
the subgroup structure of monopoles, and their relation to P-vortices.
There are strong arguements that monopoles should be associated with 
$SU(2)$ subgroups of $SU(3)$. There is also good evidence from our  $SU(3)$
lattice calculations that P-vortices pass through monopoles.  This appears
to be paradoxical at first, since P-vortices carry only center flux.
We show how these two properties can coexist (Sec.(\ref{sec_monovort})).
We begin with a general discussion of gauge-fixing 
and projection (Sec.(\ref{sec_general})).

\section{Gauges,  Maximal Gauges and Projection}
\label{sec_general}
The use of gauge-fixing followed by projection is very common in studies of
confinement.  The idea originates 
of course in 't Hooft's famous 1981 paper \cite{thooft}.
Nevertheless, it is viewed with suspicion by many physicists.  Projection
implies the deletion of a large number of gauge degrees of freedom after
choosing a particular gauge.  Can this procedure ever be a controlled
approximation?  While no definitive answer will be attempted here, we give a
discussion of some of the issues involved.

There is one very familiar example of gauge-fixing followed by projection
where the reliability of the procedure is not in question.  This is the
Coulomb gauge in non-relativistic QED.  The Coulomb gauge is a `maximal'
gauge, i.e. it seeks to squeeze the maximum physics into the timelike
sector of the gauge field.  This is accomplished by minimizing the
spatial gauge degrees of freedom.
The appropriate gauge functional is
$$
G_{coul}=\int d^{3}x  A_{k}A_{k}
$$
Minimizing $G_{coul}$ over the gauge group leads to the 
condition $\vec{\nabla}\cdot \vec{A}=0$.  
The next stage is timelike projection;
deleting the spatial components of the gauge field.
The Coulomb gauge is clearly optimal for this projection.
The truncated problem, say finding the energy levels of an atom, is then
solved non-perturbatively.  Finally, the effects of the spatial gauge field
are brought back in and treated perturbatively.

This example shows that gauge-fixing followed by projection 
can be reliable.
The key requirement 
is the presence of a small parameter.  The Coulomb gauge exploits the
fact that in atoms, electrons move slowly, with $v/c \sim \alpha$.
We can say that non-relativistic QED obeys
`timelike dominance', justified by $v/c << 1$.   Returning to 
the problem of confinement in $SU(N)$ Yang-Mills theory, no very small
parameter like the fine structure constant is expected, but moderately
small ratios can occur.   

The maximal abelian gauge (MAG) followed by
abelian projection  has been the most intensely investigated
gauge-fixing and projection scheme in non-abelian gauge theory.
For $SU(2)$, if $A_{\mu}^{3}$ is chosen as the abelian
field, then the MAG functional in the continuum is
\be
G_{mag}=\int d^{4}x[(A_{\mu}^{1})^{2}+(A_{\mu}^{2})^{2}],
\label{su2mag}
\ee
and at an extremum, the gauge conditions are
\be
\partial_{\mu}\ampm \pm i\amth \ampm =0,
\label{su2mageq}
\ee
where $\ampm =(A_{\mu}^{1} \pm iA_{\mu}^{2})/\sqrt{2}$ are the charged fields.  
This similar in spirit to
the Coulomb gauge example; to maximize the
abelian sector of the gauge field,  the charged gauge degrees of freedom 
are minimized.  After MAG gauge-fixing, abelian projection  is carried out;
the charged degrees of freedom are deleted.  
A criterion  which can give credence to 
abelian dominance is a small value for the ratio $\xi_{mon}/\xi_{\sigma}$, 
where $\xi_{mon}$ is defined by 
the size of the non-abelian core of a monopole, 
and $\xi_{\sigma}=1/\sqrt{\sigma}$ is the correlation length 
determined by the string tension, $\sigma$.
There is some evidence from lattice calculations that this ratio 
is in fact fairly small \cite{monosize}.

It is important for the generalization to $SU(3)$ to note that
an alternate way to write the $SU(2)$ MAG condition is to introduce a unit
adjoint scalar field $\vec{\Phi}$, and write
\be
G_{higgs}=
\int d^{4}x D_{\mu} \vec{\Phi} \cdot D_{\mu} \vec{\Phi}.
\label{su2higgs}
\ee
An extremum of $G_{higgs}$ is found by holding the gauge field fixed, and
varying $\vec{\Phi}$.  Then the gauge transformation that takes $\vec{\Phi}$
to the 3-axis or Higgs gauge, takes the gauge fields to the MAG.

In the $SU(3)$ work presented here, we will start from the MAG.
Although usually taken to imply a monopole/dual superconductor picture of
confinement, 
center projection is still possible; 
i.e. the `indirect' route to the 
center is open.  This gauge
 is known as the indirect maximal center gauge (IMCG).
In the IMCG, 
the relation between MAG monopoles and P-vortices can be examined
in detail.  Advocates of center dominance often recoil at the use of 
the MAG, but capturing the confining degrees of freedom in the maximal
abelian subgroup does not preclude the possibility that only the center is
really relevant.  For $SU(N)$, the center subgroup is always 
a subgroup of the maximal abelian subgroup. 

\section{Simple SU(3) MAG}
\label{sec_mag}
For $SU(3)$, there are two abelian gauge fields, $A_{\mu}^{3}$ and
$A_{\mu}^{8}$.  The remaining six gauge fields are all charged with respect
to at least one of $A_{\mu}^{3},A_{\mu}^{8}$.  Treating these charged fields
democratically, the simplest maximal abelian gauge condition for $SU(3)$ would
minimize the continuum functional
\be
G_{mag}=\int d^{4}x[(A_{\mu}^{1})^{2}+(A_{\mu}^{2})^{2}
+(A_{\mu}^{4})^{2}+(A_{\mu}^{5})^{2}+
(A_{\mu}^{6})^{2}+(A_{\mu}^{7})^{2}]
\label{su3magsimp}
\ee
over  the $SU(3)$ gauge group.  The functional \gmag is  symmetric with 
respect to each of the three $SU(2)$ subgroups of $SU(3)$.  From this it
is easy to show that at an extremum of Eq.(\ref{su3magsimp}),
there are three 
gauge conditions.  One is Eq.(\ref{su2mageq}).  The other two are
similar and involve the charged combinations
\mbox{$\ampmp=(A_{\mu}^{4} \pm i A_{\mu}^{5})/\sqrt{2}$,} and
\mbox{$\ampmpp=(A_{\mu}^{6} \pm i A_{\mu}^{7})/\sqrt{2}$}.  So with
Eq.(\ref{su3magsimp})  the $SU(3)$ MAG conditions are just $SU(2)$
MAG conditions with respect to each of the three subgroups \cite{jsu3}.

Minimizing the continuum functional \gmag is equivalent to maximizing the
lattice functional  defined by \cite{schierholz2}
\be
G_{mag}=
\sum_{x,\mu}
\left(\left|(U_{\mu})_{11}\right|^{2}+\left|(U_{\mu})_{22}\right|^{2}\right.
\left.+\left|(U_{\mu})_{33}\right|^{2}\right). 
\label{lattmag}
\ee

\begin{table}[htb]
\begin{center}
\caption{SU(3) String Tensions}
\begin{tabular}{lll}
\hline 
$\beta$ & 5.90 & 6.0\\
\hline
$SU(3)$  & 0.068(3) & 0.050(1)\\
$U(1)\times U(1)$& 0.063(3) & 0.045(2)\\
$mono$ & 0.050(2) & 0.038(1)\\
$Z(3)$ & 0.060(3) & 0.040(2)\\
\hline
\label{magtable}
\end{tabular}
\end{center}
\end{table}

We performed $SU(3)$ calculations at $\beta=5.90$ on a $10^{3}\times 16$
lattice and at $\beta=6.0$ on a $16^4$ lattice.   The details are fully
discussed elsewhere \cite{jsu3}.  The various
string tensions are summarized in Table \ref{magtable}.  
All of the cases which involve
gauge-fixing and projection made use of the MAG 
in the form of Eq.(\ref{lattmag}),
followed by abelian projection to $U(1)\times U(1)$.
There are four string tensions
listed; full $SU(3)$, MAG $U(1)\times U(1)$ , MAG monopoles, and IMCG $Z(3)$.  
The latter requires a second gauge-fixing applied to the 
abelian projected $U(1)\times U(1)$ links, followed by center projection.
The data in the table are for one gauge fixing/configuration.  
The
effect of gauge ambiguities has also been investigated, with the expected
result.  Namely, as gauge copies with higher values of the gauge functionals
are used, the projected string tensions all decrease.
For 10 copies/configuration, we
found a roughly 10\% decrease in the three 
string tensions in Table \ref{magtable} which involve gauge-fixing and
projection.

The projected string tensions 
of Table \ref{magtable} are all smaller than the corresponding full
$SU(3)$ values. 
This behavior is quite different from that found in $SU(2)$.
For $SU(2)$, 
the MAG $U(1)$ string tension is {\it larger} than the $SU(2)$ string tension
for one gauge copy/configuration.  Taking gauge copies with higher functional
values causes the $U(1)$ string tension to decrease and approach the
full $SU(2)$ string tension.  For $SU(3)$, the abelian projected $U(1) \times
U(1)$ string tension is already low for one gauge copy/configuration.  Taking
account of `more maximal' gauge copies reduces it still further below the
$SU(3)$ string tension.  The low results found for MAG monopole and IMCG 
$Z(3)$ string tensions are certainly caused in part by this difficulty.
We conclude that either something is wrong with the whole
idea of the MAG and abelian projection for 
$SU(3)$, or that a different form of the MAG is
needed.  In the next section, we explore the latter possibility.

\section{Generalized Maximal Abelian Gauge}
\label{sec_mag_gen}
For $SU(2)$ the functional $G_{higgs}$ of Eq.(\ref{su2higgs})  is merely a
rewriting of
the MAG functional, Eq.(\ref{su2mag}).  However, the analogous
statement is not true 
for $SU(3)$.  
Introduce a unit length adjoint scalar field for $SU(3)$.  In Higgs gauge,
this field 
is gauge transformed to the
Cartan algebra,
$$ 
\vec{\Phi}=\hat{3}\cos(\chi)+\hat{8}\sin(\chi).
$$
where the angle $\chi$ is $SU(3)$ gauge invariant.
In  this gauge,  the $SU(3)$ version of the functional of
Eq.(\ref{su2higgs}) 
becomes
\be
G_{higgs}  =  \int d^{4}x
\left\{ \right.
\cos^{2}(\chi)\left[(A_{\mu}^{1})^{2}+(A_{\mu}^{2})^{2}\right]+
\label{su3maghiggs}
\ee
$$
\cos^{2}(\frac{\pi}{3}-\chi)\left[(A_{\mu}^{4})^{2}+(A_{\mu}^{5})^{2}\right]
+\cos^{2}
(\frac{\pi}{3}+\chi)\left[(A_{\mu}^{6})^{2}\right.
\left.\left.+(A_{\mu}^{7})^{2}\right]\right\},
$$
which is clearly different from Eq.(\ref{su3magsimp}).  A single $SU(3)$ 
adjoint
scalar field cannot give equal coefficients to the charged fields in the
different $SU(2)$ subgroups.
Formally, equal coefficients can be attained by averaging over the
angle $\chi$, but this is in effect saying there is a continuous 
distribution of scalar fields with different $\chi$ angles.  A continuous
distribution is actually unnecessary; it suffices to have two 
fields, at angles $\chi$ and $\chi + \pi/2$ relative to $\hat{3}$.
This way of obtaining the MAG functional of Eq.(\ref{su3magsimp})  suggests 
that  Eq.(\ref{su3magsimp}) is itself rather unnatural. 
Minimizing this functional involves attempting 
to suppress the charged gauge fields with respect to (at least) two
different directions in the Cartan algebra.  These conflicting requirements
may be the cause of the low string tensions found in 
Table \ref{magtable}.

The functional of 
Eq.(\ref{su3maghiggs}) can still be regarded as defining a maximal abelian gauge
since minimizing it will tend to suppress all the charged gauge fields.
The equivalent lattice functional  
is easily written down;
\be
G_{higgs}
=\sum_{x,\mu} {\rm tr}((\lambda_{3}\cos\chi+\lambda_{8}\sin\chi)U_{\mu}(x)
\left.(\lambda_{3}\cos\chi^{\prime}+\lambda_{8}\sin\chi^{\prime})U_{\mu}^{\dagger}(x)\right),
\label{latthiggs}
\ee
where $\chi=\chi(x)$ and $\chi^{\prime}=\chi(x+\hat{\mu}a)$. 
Eq.(\ref{latthiggs}) allows two distinct possibilities for a generalized MAG.
The angle $\chi$ can be held fixed, or allowed to
vary with $x$.  
(The latter possibility  requires an extra term,
$\partial_{\mu}\chi \partial_{\mu}\chi$,  in the integrand of
Eq.(\ref{su3maghiggs}).)
Either way, there is now just  
one effective Higgs field. 
We are at present actively exploring 
Eq.(\ref{latthiggs}) both for the case where $\chi$ is held fixed and
where it is allowed to vary.  Whether $\chi$ is held fixed or allowed to vary,
the  functional describes 
a `maximal' gauge, so there will be gauge ambiguities.  An
inexpensive way to see if Eq.(\ref{latthiggs}) leads to an improved abelian
projection will be to calculate with one gauge-fixing/configuration.  
The result
should be a $U(1) \times U(1)$ string tension {\it larger} than the full
$SU(3)$ result, which would then decrease when gauge copies with higher
functional values are used.

The overall 
message of this section is that Eq.(\ref{lattmag}) is based on a too-literal
analogy with the $SU(2)$ case, and that more general possibilities for the 
$SU(3)$ MAG exist and are needed.

\section{Monopoles and P-Vortices in SU(3)}
\label{sec_monovort}
There are interesting questions about monopoles and vortices and their
relation to the degrees of freedom which control confinement.  
For example, how are the monopoles and P-vortices found  on the 
lattice related to physical monopoles and center vortices?  
Is either of these degrees of freedom more fundamental than the other, 
and if so which?  In this section, we discuss a question 
which first arises upon turning to an $SU(3)$ gauge group from $SU(2)$.

 In an abelian or Higgs gauge,
outside the core of a monopole, its fields are abelian.  For an
$SU(2)$ gauge group, this means the long range gauge field is 
$\sim A_{\mu}^{3} \tau_{3}/2$, where $\tau_{3}$ is a Pauli matrix.
For an $SU(3)$ gauge group,  a generic abelian field is of the form
$\sim (A_{\mu}^{3} \lambda_{3}+A_{\mu}^{8} \lambda_{8})/2$.
In his original paper, `t Hooft argued that for $SU(N)$, 
monopoles should be associated with $SU(2)$ subgroups \cite{thooft}.
Studying monopole solutions, E. Weinberg and P. Yi came to
the same conclusion \cite{eweinberg}.  These results imply that in $SU(3)$,
the long range field of a monopole is more specific than the generic
abelian field.  Namely, the field should 
always be `$\lambda_{3}$-like', and there should be no
purely `$\lambda_{8}$-like' monopoles.  We find support for this
in our $SU(3)$ lattice calculations.  The color magnetic current we find is
generally of the $\lambda_{3}$-like forms (1,-1,0), (1,0,-1), (0,1,-1).
The $\lambda_{8}$-like forms (1,1,-2), (1,-2,1), (-2,1,1) basically
never happen.  

If it is granted that monopoles are associated with $SU(2)$ subgroups, it 
is still possible that the net effect of 
superposing monopoles from different subgroups could be fields which
mainly connect to the group center. 
However, from one monopole, the field is alive in only two  out of three
colors, whereas center flux is the same for all three.   This would seem
to argue against a detailed connection on the lattice between
MAG monopoles and IMCG P-vortices.

It is well-established in $SU(2)$ lattice calculations
that in the IMCG, P-vortices pass through 
MAG monopoles \cite{greensitemono,stacktucker}.  This can 
be pictured as a squeezing of the monopole flux into $Z(2)$ Dirac strings.
For $SU(2)$, the monopole flux $g$ comes in Schwinger units, $g=4\pi/e$, where
$e$ is the non-abelian gauge coupling.  If
the  monopole flux is squeezed into two strings, each carrying flux $g/2$, the
strings are visible to fundamental (I=1/2) quarks, and behave like 
P-vortices.  

In our $SU(3)$ lattice calculations, using MAG monopoles and
IMCG \mbox{P-vortices,} it was found that again there is an intimate connection
between monopole and vortex degrees of freedom.  If a link has a non-zero
magnetic current, the cube dual to this link has faces pierced by the
$Z(3)$ flux of P-vortices over 80\% of the time.  Since methods of locating
monopoles and P-vortices are not precise, the actual number could be 100\%.
The $SU(3)$ case allows 
an odd number of  P-vortices to meet at a monopole, but otherwise
there appears to be little essential difference 
in lattice numerical calculations between $SU(2)$ and $SU(3)$.

To see how $SU(3)$ P-vortices carrying only center flux can fit together with
monopoles associated with $SU(2)$ subgroups, consider a specific example.
Put a monopole at the origin, and let it be associated with 
the `I-spin' subgroup generated by  
$\lambda_{1},\lambda_{2},\lambda_{3}$.  Imitate the $SU(2)$
situation by squeezing the monopole
 flux into two strings on the $z$-axis, one on
the $+z$-axis carrying upward flux $g/2$; the other on the $-z$-axis
carrying upward flux $-g/2$.  Now suppose 
that there is an $SU(3)$ P-vortex on the
$z$-axis.  A P-vortex has flux quantized in units $\Phi_{P}=g/\sqrt{3}$.
For $z>0$, we can represent the vortex in terms of the $\lambda_{8}$-like matrix
associated with $V$-spin, 
\mbox{$\lambda_{8}^{\prime}=(-\lambda_{8}+\sqrt{3} \lambda_{3})/2$.}
When a fundamental quark goes around this part of the P-vortex, it picks up a 
phase factor
$$\exp(ie\Phi_{P}\frac{\lambda_{8}^{\prime}}{2}) \subset Z(3).$$ 
For $z<0$, we can represent the vortex in
terms of the $\lambda_{8}$-like matrix associated with $U$-spin,
$\lambda_{8}^{\prime\prime}=(-\lambda_{8}-\sqrt{3} \lambda_{3})/2$, giving
the {\it same} phase factor
$$\exp(ie\Phi_{P}\frac{\lambda_{8}^{\prime\prime}}{2}) \subset Z(3).$$ 
We note that for $z>0$ the part of the flux involving $\lambda_{3}/2$ 
is $\Phi_{P}\sqrt{3}/2=g/2$, while that for $z<0$ involving $\lambda_{3}/2$
is $-\Phi_{P}\sqrt{3}/2=-g/2$. These are 
exactly as they should be for an I-spin monopole at the origin.  The flux
of the P-vortex has a different distribution over colors on opposite sides of
the monopole.  This flux difference is not visible in the phase factor
experienced by a quark encircling the P-vortex.  All this is entirely
analogous to the $SU(2)$ situation.

We conclude that in $SU(3)$ 
there is no conflict between the following three facts:
(1) Monopoles are associated with $SU(2)$ subgroups.  (2) P-vortices carry
only center flux.  (3) The magnetic current of monopoles is  a world line
located on the world sheet of P-vortices.  The question of  
which  of MAG monopoles or IMCG P-vortices is most closely
tied to physical topological objects remains unanswered.

\end{document}